\documentclass[12pt]{article}
\usepackage{amsmath,amssymb,amsthm,amsxtra,overpic,bbm,bm,epsfig,subfigure}
\usepackage{mathrsfs}
\usepackage{graphicx}
\usepackage{color}
\usepackage{comment}
\usepackage{float}
\usepackage{cite}
\usepackage{bbold}
\usepackage{slashed,stmaryrd}

\def\thefootnote{\fnsymbol{footnote}}

\begin{document}

\vspace{0.2cm}

\begin{center}
{\Large\bf Transition Probabilities for Flavor Eigenstates of Non-Hermitian Hamiltonians in the PT-Broken Phase}
\end{center}

\vspace{0.2cm}

\begin{center}
{\bf Tommy Ohlsson}~$^{a,~b,~c}$~\footnote{E-mail: tohlsson@kth.se},
\quad
{\bf Shun Zhou}~$^{d,~e}$~\footnote{E-mail: zhoush@ihep.ac.cn}
\\
\vspace{0.2cm}
{\small $^a$Department of Physics, School of Engineering Sciences, KTH Royal Institute of Technology, \\
AlbaNova University Center, Roslagstullsbacken 21, SE-106 91 Stockholm, Sweden \\
$^b$The Oskar Klein Centre for Cosmoparticle Physics, AlbaNova University Center, \\
Roslagstullsbacken 21, SE-106 91 Stockholm, Sweden \\
$^c$University of Iceland, Science Institute, Dunhaga 3, IS-107  Reykjavik,  Iceland\\
$^d$Institute of High Energy Physics, Chinese Academy of Sciences, Beijing 100049, China\\
$^e$School of Physical Sciences, University of Chinese Academy of Sciences, Beijing 100049, China}
\end{center}

\vspace{1.5cm}

\begin{abstract}
We investigate the transition probabilities for the ``flavor" eigenstates in the two-level quantum system, which is described by a non-Hermitian Hamiltonian with the parity and time-reversal (PT) symmetry. Particularly, we concentrate on the so-called PT-broken phase, where two eigenvalues of the non-Hermitian Hamiltonian turn out to be a complex conjugate pair. In this case, we find that the transition probabilities will be unbounded in the limit of infinite time $t \to +\infty$. However, after performing a connection between a non-Hermitian system, which exhibits passive PT-symmetry and global decay, and the neutral-meson system in particle physics, we observe that the diverging behavior of the transition probabilities is actually applicable to the gauge-transformed neutral-meson states, whereas the transition probabilities for physical states are exponentially suppressed by the global decay. We also present a brief review on the situation at the so-called exceptional point, where both the eigenvalues and eigenvectors of the Hamiltonian coalesce.
\end{abstract}

\newpage

\def\thefootnote{\arabic{footnote}}
\setcounter{footnote}{0}

\section{Introduction}

Non-Hermitian Hamiltonians with the joint parity and time-reversal (PT) symmetry have recently attracted a lot of attention~\cite{Bender:2007nj} and very interesting applications have been found for a number of physical systems in particle physics, nuclear physics, optics, electronics, and many others~\cite{review}. In the existing literature, the two-level system with PT-symmetric non-Hermitian Hamiltonians~\cite{Bender:2002yp,Bender:2002vv, Feng:2017, Joglekar:2018, Wang:2019, Roccati:2021} has been extensively investigated, as a simple but instructive example, to explore and clarify all related conceptual issues~\cite{Mostafazadeh:2008pw, Kleefeld:2009vd, Mannheim:2009zj}. However, it is worthwhile to mention that the transition amplitudes and probabilities between the ``flavor" eigenstates have rarely been studied, except for some general discussion in Refs.~\cite{Bagarello1, Bagarello2}.

Since the phenomenon of ``flavor" mixing is quite common in particle physics, such as flavor oscillations of massive neutrinos~\cite{Xing:2019vks} and the neutral-meson system $P^0$-$\overline{P}^0$~\cite{Tanabashi:2018oca} (e.g., $K^0$-$\overline{K}^0$, $D^0$-$\overline{D}^0$, and $B^0$-$\overline{B}^0$), it is intriguing to consider the transitions among ``flavor" eigenstates in the system with PT-symmetric non-Hermitian Hamiltonians~\cite{Ohlsson:2015xsa, Ohlsson:2019noy}. In our previous work~\cite{Ohlsson:2019noy}, we calculated the transition probabilities for the ``flavor" eigenstates in the scenario, where the PT symmetry is always preserved and two eigenvalues of the non-Hermitian Hamiltonian are real, which is known as the \emph{PT-symmetric phase}. In the present work, we aim to extend the previous study in the PT-symmetric phase to the \emph{PT-broken phase}. The primary motivation for such an extension is at least two-fold.

First, by ``flavor" eigenstates of a two-level quantum system with a non-Hermitian Hamiltonian ${\cal H}$, we mean the complete set of basis vectors $\{|u^{}_\beta\rangle\}$ (for $\beta = a, b$), in which the matrix representation of the Hamiltonian is given by ${\cal H}^{}_{\alpha \beta} \equiv \langle u^{}_\alpha|{\cal H}|u^{}_\beta\rangle$ and the left vectors $\langle u^{}_\alpha| \equiv |u^{}_\alpha\rangle^\dagger$ and $\langle u^{}_\beta| \equiv |u^{}_\beta\rangle^\dagger$ (for $\alpha, \beta = a, b$) have been defined as in conventional quantum mechanics. Therefore, the transition amplitudes in the following discussion will be referred to the projection of the time-evolved flavor eigenstates $\{|u^{}_\alpha(t)\rangle\}$ into their initial states $\{|u^{}_\beta\rangle\}$. However, whenever the transition amplitudes ${\cal A}^{}_{\alpha \beta} \equiv \langle u^{}_\beta|u^{}_\alpha(t)\rangle$ are calculated, we will clearly indicate the exact definition of the involved inner product as well as that of the left state vectors. As mentioned, for PT-symmetric non-Hermitian Hamiltonians, the transition amplitudes ${\cal A}^{}_{\alpha \beta}$ and the corresponding probabilities ${\cal P}^{}_{\alpha \beta} \equiv |{\cal A}^{}_{\alpha \beta}|^2$ have been explicitly computed and extensively studied in Ref.~\cite{Ohlsson:2019noy}. Hence, it is a natural continuation to extend the investigation to the PT-broken phase.

Second, in contrast to the PT-symmetric phase, where two eigenvalues of the non-Hermitian Hamiltonian are real, the PT-broken phase will be complicated by a complex-conjugate pair of eigenvalues. As is well known, if PT symmetry is maintained, it is always possible to find a similarity transformation that converts a non-Hermitian Hamiltonian into its Hermitian counterpart~\cite{Mostafazadeh:2008pw, Mannheim:2009zj, Ohlsson:2019noy}. However, this is impossible for the PT-broken phase, rendering it rather different. Hence, in this particular case, the transition probabilities deserve a dedicated study.

The remaining part of this work is organized as follows. In Sec.~\ref{sec: general}, we present the general formalism for the investigation of PT-symmetric non-Hermitian Hamiltonians and summarize the main features of the PT-symmetric phase, the PT-broken phase, and the \emph{exceptional point}, where the transition between these two phases occurs. Then, in Sec.~\ref{sec: broken}, the transition amplitudes and probabilities in the PT-broken phase will be introduced and studied, where the connection between the PT-broken phase and the neutral-meson system is also performed. Finally, in Sec.~\ref{sec: conclusion}, we summarize the main results and draw our conclusions.

\section{General Formalism}\label{sec: general}

For a general discussion about the properties of PT-symmetric non-Hermitian Hamiltonians and their applications, one should be referred to the excellent review by Bender~\cite{Bender:2007nj} and references therein. Particularly, in this work, we focus on the simple two-level system, for which the Hamiltonian is diagonalizable and space-time independent. The space-reflection operator ${\cal P}$ is defined as~\cite{Ohlsson:2019noy}
\begin{equation}
{\cal P} = \left(\begin{matrix} 0 & 1 \cr 1 & 0\end{matrix}\right) \; ,
\end{equation}
and the time-reversal operator ${\cal T}$ is taken to be just the complex conjugation ${\cal K}$, namely, ${\cal T} {\cal O} {\cal T}^{-1} = {\cal O}^*$ for any operators ${\cal O}$ in the Hilbert space. The most general form of the Hamiltonian for the two-level system is given by
\begin{equation}
{\cal H} = \left(\begin{matrix} a & b \cr c & d\end{matrix}\right) \; ,
\end{equation}
where $\{a, b, c, d\}$ are arbitrary complex constants. The PT symmetry of the Hamiltonian system requires that $[{\cal PT}, {\cal H}] = {\bf 0}$, so we have
\begin{eqnarray}
\left({\cal PT} {\cal H}\right) \Psi &=& \left(\begin{matrix} 0 & 1 \cr 1 & 0\end{matrix}\right) \left(\begin{matrix} a^* & b^* \cr c^* & d^*\end{matrix}\right) \Psi^* = \left(\begin{matrix} c^* & d^* \cr a^* & b^*\end{matrix}\right) \Psi^*\; , \label{eq:PTH}\\
\left({\cal H}{\cal PT}\right) \Psi &=& \left(\begin{matrix} a & b \cr c & d\end{matrix}\right) \left(\begin{matrix} 0 & 1 \cr 1 & 0\end{matrix}\right) \Psi^* = \left(\begin{matrix} b & a \cr d & c\end{matrix}\right) \Psi^* \; , \label{eq:HPT}
\end{eqnarray}
where $\Psi$ stands for any vectors in the Hilbert space that the operators are acting on. From Eqs. (\ref{eq:PTH}) and (\ref{eq:HPT}), one can recognize that the PT symmetry of the system implies that $a = d^*$ and $b = c^*$. Therefore, the most general non-Hermitian Hamiltonian ${\cal H}$ actually contains only four degrees of freedom (in terms of the number of real parameters), when it respects the PT symmetry. This is equal to the number of free parameters in the two-level system with the Hermitian Hamiltonian, where $a$ and $d$ are real while $b = c^*$.

\begin{figure}[!t]
\begin{center}
\includegraphics[width=0.6\textwidth]{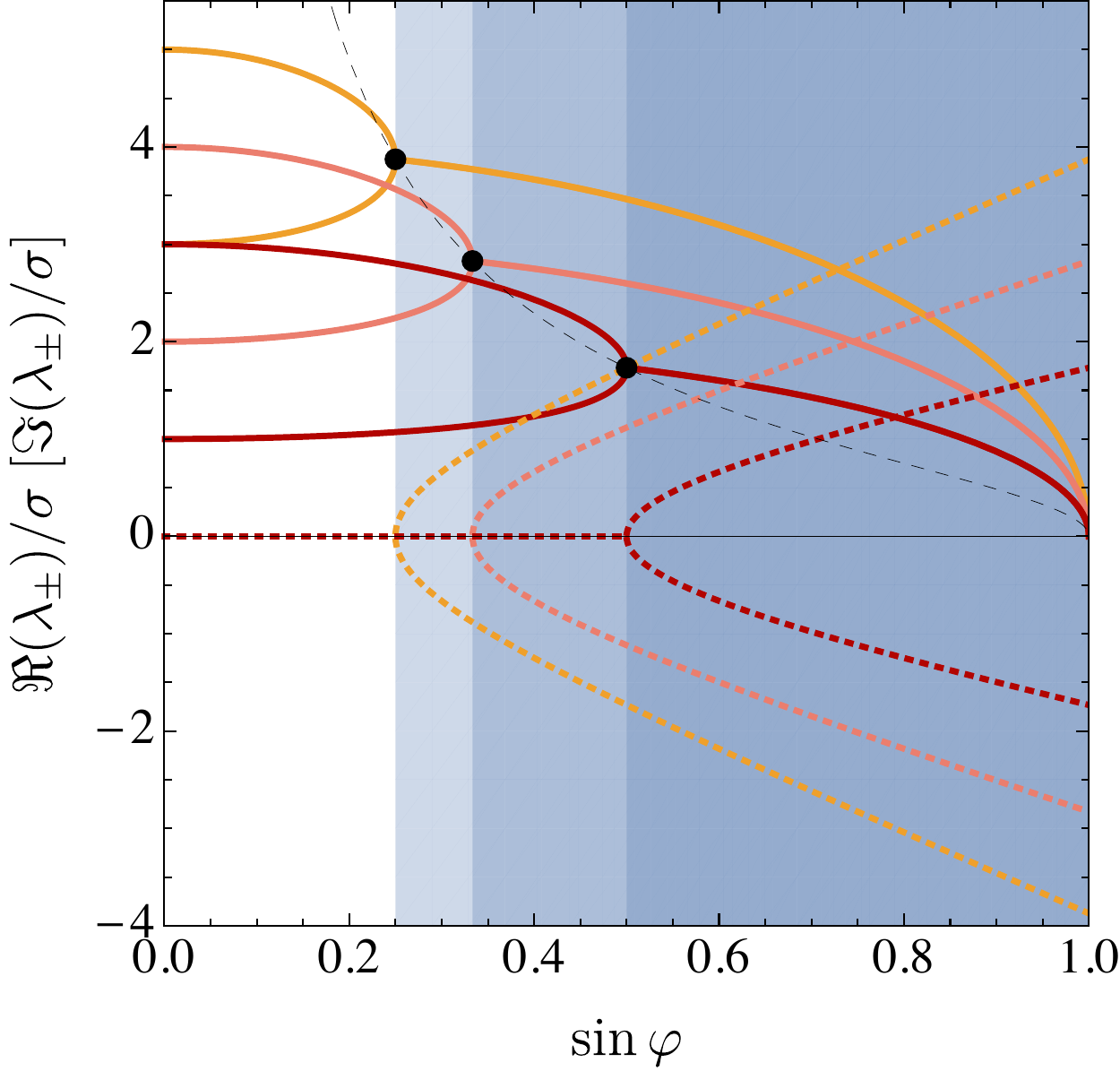}
\end{center}
\vspace{-0.6cm}
\caption{Illustration for the real (solid curves) and imaginary (dotted curves) parts of the normalized eigenvalues $\lambda_\pm/\sigma =  \xi \sqrt{1- \sin^2 \varphi} \pm \sqrt{1- \xi^2 \sin^2\varphi}$ as functions of $\sin \varphi$ for three different choices of the ratio of the parameters $\rho$ and $\sigma$, i.e., $\xi \equiv \rho/\sigma = 2~\mbox{(red curves)}, 3~\mbox{(orange curves)}, 4~\mbox{(yellow curves)}$, in the \emph{PT-symmetric phase} (none or less shaded areas) and the \emph{PT-broken phase} (shaded areas). The corresponding \emph{exceptional points} are marked by black points (`$\bullet$') and the black thin dashed curve shows the trajectory of the exceptional points.}
\label{fig:plotEigenvalues}
\end{figure}

For later convenience, we adopt the following parametrization of the most general PT-symmetric non-Hermitian Hamiltonian, viz.,
\begin{equation}
{\cal H} = \left(\begin{matrix} \rho e^{+{\rm i}\varphi} & \sigma e^{+{\rm i}\phi} \cr  \sigma e^{-{\rm i}\phi} & \rho e^{-{\rm i}\varphi} \end{matrix}\right) \; , \label{eq:H}
\end{equation}
where all parameters $\{\rho, \varphi\}$ and $\{\sigma, \phi\}$ are real and time-independent. For a recent study on time-dependent parameters in ${\cal H}$, see Refs.~\cite{Grimaudo:2019zli,Ju:2019kso}. With the Hamiltonian in Eq.~(\ref{eq:H}), one can immediately figure out its two eigenvalues
\begin{equation}
\lambda^{}_\pm = \rho \cos \varphi \pm \sqrt{\sigma^2 - \rho^2 \sin^2\varphi} \; .
\label{eq:lambdaPTsymm}
\end{equation}
Under the condition that $\rho^2\sin^2\varphi < \sigma^2$ is satisfied, the two eigenvalues are real. If this condition is not satisfied, namely, $\rho^2 \sin^2\varphi \geq \sigma^2$, we obtain either (i) two complex eigenvalues (if $\rho^2 \sin^2\varphi > \sigma^2$ holds)
\begin{equation}
\lambda_\pm = \rho \cos \varphi \pm {\rm i} \sqrt{\rho^2 \sin^2\varphi - \sigma^2} \; ,
\label{eq:lambdaPTbroken}
\end{equation}
which are complex conjugates to each other, or (ii) a degenerate real eigenvalue $\lambda_\pm = \lambda_0 = \rho \cos \varphi$ with multiplicity 2, since $\rho^2 \sin^2\varphi = \sigma^2$ holds. In Fig.~\ref{fig:plotEigenvalues}, we present the eigenvalues as a function of $\sin \varphi$ for different choices of the ratio of the parameters $\rho$ and $\sigma$.
The eigenvalues are displayed in the PT-symmetric phase ($\rho^2 \sin^2\varphi < \sigma^2$: two real eigenvalues, see Subsec.~\ref{sub:PTsymmetric}) and the PT-broken phase ($\rho^2 \sin^2\varphi > \sigma^2$: two complex-conjugate eigenvalues, see Subsec.~\ref{sub:PTbroken}) as well as the exceptional points are indicated ($\rho^2 \sin^2\varphi = \sigma^2$: one degenerate real eigenvalue, see Subsec.~\ref{sub:EP}). Some helpful comments on the eigenvalues and their corresponding eigenvectors of the non-Hermitian Hamiltonian ${\cal H}$ in Eq.~(\ref{eq:H}) are in order.

\subsection{PT-Symmetric Phase}
\label{sub:PTsymmetric}

Although the PT-symmetric phase has been carefully studied in Ref.~\cite{Ohlsson:2019noy}, we briefly summarize its main features in this subsection in order to make the presentation self-consistent, to clarify our notation, and to easily perform a direct comparison with the PT-broken phase and the exceptional point in the next subsections. As mentioned before, the two eigenvalues in Eq.~(\ref{eq:lambdaPTsymm}) are real if the condition $\rho^2 \sin^2\varphi < \sigma^2$ is fulfilled. This is usually called the PT-symmetric phase of the system. In this case, we write the two eigenvectors corresponding to $E^{}_\pm = \lambda_\pm$ as
\begin{equation}
|u^{}_+\rangle = \frac{1}{\sqrt{2\sin \alpha}}\left(\begin{matrix} e^{+{\rm i}\pi/4} \cdot e^{-{\rm i}\alpha/2} \cr e^{-{\rm i}\pi/4} \cdot e^{+{\rm i}\alpha/2}\end{matrix}\right) \;, \quad |u^{}_-\rangle = \frac{\rm i}{\sqrt{2\sin\alpha}} \left(\begin{matrix} e^{+{\rm i}\pi/4} \cdot e^{+{\rm i}\alpha/2} \cr e^{-{\rm i}\pi/4} \cdot e^{-{\rm i}\alpha/2}\end{matrix}\right) \; , \label{eq:u+u-2}
\end{equation}
where $\cos \alpha \equiv (\rho \sin\varphi)/\sigma$ has been defined and $\phi = 0$ is assumed.\footnote{Note that the definition of the parameter $\alpha$ differs from that in Ref.~\cite{Ohlsson:2019noy}, where $\sin\alpha \equiv (\rho\sin\varphi)/\sigma$. The reason for such a change is to make a coherent presentation in both the PT-symmetric and PT-broken phases.} One can easily verify that ${\cal PT}|u^{}_\pm\rangle = \pm |u^{}_\pm\rangle$, indicating that $|u^{}_\pm\rangle$ are also the eigenvectors of the ${\cal PT}$ operator.

On the other hand, we can apply the bi-orthogonal formalism~\cite{Sachs1963} to the non-Hermitian Hamiltonian in Eq.~(\ref{eq:H}) but with $\phi = 0$ assumed for illustration, and then introduce the so-called metric operator $\eta$~\cite{Mostafazadeh:2008pw}, i.e.,
\begin{equation}
\eta \equiv \sum_{s = \pm} |v^{}_s\rangle \langle v^{}_s| = {\cal P} \left(|u^{}_+\rangle \langle u^{}_+| + |u^{}_- \rangle \langle u^{}_-|\right) {\cal P} = \left(\begin{matrix} \csc\alpha & -{\rm i}\cot \alpha \cr +{\rm i}\cot\alpha & \csc\alpha \end{matrix}\right) \; ,
\end{equation}
where $\langle v^{}_\pm| \equiv |v^{}_\pm\rangle^\dagger$ and $\langle u^{}_\pm| \equiv |u^{}_\pm\rangle^\dagger$. By construction, the relation $\eta {\cal H} \eta^{-1} = {\cal H}^\dagger$ holds, so one can easily prove that there exists a charge-conjugation operator ${\cal C}$ defined as \cite{Bender:2004zz}
\begin{equation}
{\cal C} \equiv {\cal P}^{-1} \eta = \eta^{-1} {\cal P} = \left(\begin{matrix} +{\rm i}\cot\alpha & \csc\alpha \cr \csc\alpha & -{\rm i}\cot\alpha \end{matrix}\right) \; , \label{eq:C}
\end{equation}
satisfying the commutation relation $[{\cal C}, {\cal H}] = {\bf 0}$. This non-Hermitian Hamiltonian system respects both the ${\cal C}$ and ${\cal PT}$ symmetries, and thus, the ${\cal CPT}$ symmetry. Since the ${\cal PT}$-inner product is actually not positive-definite (due to ${\rm det}~{\cal P} = -1 < 0$), it is necessary to introduce the $\eta$- and ${\cal CPT}$-inner products for the definitions of the transition amplitudes and probabilities \cite{Mannheim:2017apd}. More explicitly,
\begin{itemize}
\item {\it The $\eta$-inner product} for any two state vectors $|\psi\rangle$ and $|\chi\rangle$ reads
    \begin{equation}
    \langle \psi|\chi \rangle^{}_\eta \equiv \langle \psi| \eta |\chi \rangle = |\psi\rangle^\dagger \cdot \eta \cdot |\chi \rangle \; .
    \end{equation}

\item {\it The ${\cal CPT}$-inner product} for any two state vectors $|\psi\rangle$ and $|\chi\rangle$ reads
        \begin{equation}
    \langle \psi|\chi \rangle^{}_{\cal CPT} \equiv \left({\cal CPT}|\psi\rangle\right)^{\rm T} \cdot |\chi \rangle = |\psi\rangle^\dagger \cdot {\cal PC} \cdot |\chi \rangle = \langle \psi|\chi \rangle^{}_\eta \; , \label{eq:CPT}
    \end{equation}
    where ${\cal PC} = \eta$ from the definition of the ${\cal C}$ operator in Eq.~(\ref{eq:C}) has been used in the last step in Eq.~(\ref{eq:CPT}).
\end{itemize}
Therefore, the $\eta$- and ${\cal CPT}$-inner products are equivalent and we can use either of them to calculate the transition amplitudes and probabilities between two quantum states. These calculations have been performed in Ref.~\cite{Ohlsson:2019noy}.

\subsection{PT-Broken Phase}
\label{sub:PTbroken}

Under the condition $\sigma^2 < \rho^2\sin^2\varphi$, one can check that $[{\cal PT}, {\cal H}] = {\bf 0}$ remains to be valid for the most general form of the Hamiltonian ${\cal H}$ in Eq.~(\ref{eq:H}). This should be the case as we have derived the general form of the PT-symmetric Hamiltonian for arbitrary values of the parameters. However, as we have shown in Eq.~(\ref{eq:lambdaPTbroken}), the Hamiltonian~(\ref{eq:H}) has two complex eigenvalues $E^\prime_\pm = \lambda_\pm$, which in this PT-broken phase are labeled by primes in order to avoid any confusion with the ones in the PT-symmetric phase.

Following the same procedure as in the PT-symmetric phase to calculate the eigenvectors $|u^\prime_\pm\rangle \equiv (a^\prime_\pm, b^\prime_\pm)^{\rm T}$ corresponding to the eigenvalues $E^\prime_\pm$, we obtain
\begin{equation}
|u^\prime_+\rangle = N^\prime_+ \left(\begin{matrix} e^{+{\rm i}\pi/4} \cdot e^{+(\alpha^\prime + {\rm i}\phi)/2} \cr e^{-{\rm i}\pi/4} \cdot e^{-(\alpha^\prime + {\rm i}\phi)/2} \end{matrix}\right) \; , \quad |u^\prime_-\rangle = N^\prime_- \left(\begin{matrix} e^{+{\rm i}\pi/4} \cdot e^{-(\alpha^\prime - {\rm i}\phi)/2} \cr e^{-{\rm i}\pi/4} \cdot e^{+(\alpha^\prime - {\rm i}\phi)/2} \end{matrix}\right) \label{eq:eigenvectors_bp}
\end{equation}
where $\cosh\alpha^\prime \equiv (\rho \sin\varphi)/\sigma$ has been defined. Performing a comparison between Eq.~(\ref{eq:u+u-2}) in the PT-symmetric phase and Eq.~(\ref{eq:eigenvectors_bp}) in the PT-broken phase, one observes the connection between these two cases by simply identifying $\alpha^\prime = -{\rm i}\alpha$ up to the normalization constants and the phase $\phi$. Then, we try to fix the two normalization constants $N^\prime_\pm$ by examining the ${\cal PT}$-inner products of these two eigenvectors, namely,
\begin{eqnarray}
\langle u^\prime_+| u^\prime_+\rangle^{}_{\cal PT} &=& - 2|N^\prime_+|^2 \sin\phi \; , \quad \langle u^\prime_-| u^\prime_-\rangle^{}_{\cal PT} = - 2|N^\prime_-|^2 \sin\phi \; , \label{eq:u+u+u-u-}\\
\langle u^\prime_+| u^\prime_-\rangle^{}_{\cal PT} &=& -2 {\rm i} N^{\prime *}_+  N^\prime_- \sinh (\alpha^\prime - {\rm i}\phi) \; , \quad \langle u^\prime_-| u^\prime_+\rangle^{}_{\cal PT} = +2 {\rm i} N^{\prime *}_-  N^\prime_+ \sinh (\alpha^\prime + {\rm i}\phi) \; . \label{eq:u+u-u-u+}
\end{eqnarray}
At first sight, it seems that one can choose proper values of $N^\prime_\pm$ to guarantee the orthogonality conditions $\langle u^\prime_\pm | u^\prime_\mp \rangle^{}_{\cal PT} = 0$. However, as one can observe from Eq.~(\ref{eq:u+u-u-u+}), this is only possible if both $\alpha^\prime = 0$ and $\sin\phi = 0$ hold, or equivalently, at the so-called exceptional point with $\rho^2 \sin^2\varphi = \sigma^2$.

Therefore, for $\rho^2 \sin^2\varphi > \sigma^2$ under consideration, we have to determine the normalization constants $N^\prime_\pm$ by requiring $\langle u^\prime_\pm | u^\prime_\pm \rangle^{}_{\cal PT} = 0$ and $\langle u^\prime_\pm| u^\prime_\mp\rangle^{}_{\cal PT} = + 1$. These requirements differ significantly from those in the PT-symmetric phase. From Eq.~(\ref{eq:u+u+u-u-}) with $\langle u^\prime_\pm | u^\prime_\pm \rangle^{}_{\cal PT} = 0$, we immediately get $\phi = 0$ (or $\phi = \pi$), which is also consistent with our previous convention in the PT-symmetric phase. In addition, from Eq.~(\ref{eq:u+u-u-u+}) with $\phi = 0$ and $\langle u^\prime_\pm| u^\prime_\mp\rangle^{}_{\cal PT} = + 1$, we obtain $N^\prime_+ = e^{-{\rm i}\pi/4}/\sqrt{2\sinh \alpha^\prime}$ and $N^\prime_- = e^{+{\rm i}\pi/4}/\sqrt{2\sinh \alpha^\prime}$, and thus, using Eq.~(\ref{eq:eigenvectors_bp}), we find the two eigenvectors as
\begin{eqnarray}
|u^\prime_+\rangle &=& \frac{e^{-{\rm i}\pi/4}}{\sqrt{2\sinh \alpha^\prime}} \left(\begin{matrix} e^{+{\rm i}\pi/4} \cdot e^{+\alpha^\prime/2} \cr e^{-{\rm i}\pi/4} \cdot e^{-\alpha^\prime/2} \end{matrix}\right) = \frac{1}{\sqrt{2\sinh \alpha^\prime}} \left(\begin{matrix} e^{+\alpha^\prime/2} \cr -{\rm i} e^{-\alpha^\prime/2} \end{matrix}\right) \; , \label{eq:up+}\\
|u^\prime_-\rangle &=& \frac{e^{+{\rm i}\pi/4}}{\sqrt{2\sinh \alpha^\prime}} \left(\begin{matrix} e^{+{\rm i}\pi/4} \cdot e^{-\alpha^\prime/2} \cr e^{-{\rm i}\pi/4} \cdot e^{+\alpha^\prime/2} \end{matrix}\right) = \frac{1}{\sqrt{2\sinh \alpha^\prime}}  \left(\begin{matrix} +{\rm i} e^{-\alpha^\prime/2} \cr  e^{+\alpha^\prime/2} \end{matrix}\right)\; . \label{eq:up-}
\end{eqnarray}
It is helpful to make some comments on the further connection between the PT-symmetric and PT-broken phases. In the latter case, we have two eigenvalues $E^\prime_\pm = \rho \cos\varphi \pm {\rm i} \sigma \sinh \alpha^\prime$, in which the replacement of $\alpha^\prime = -{\rm i}\alpha$ leads to the two eigenvalues $E^{}_\pm$ in the former case. At the same time, if we replace $\alpha^\prime$ by $-{\rm i}\alpha$ everywhere in Eqs.~(\ref{eq:up+}) and (\ref{eq:up-}), the eigenvectors $|u^\prime_\pm \rangle$ will reduce to $|u^{}_\pm\rangle$ in Eq.~(\ref{eq:u+u-2}).

Nevertheless, given the eigenvectors $|u^\prime_\pm\rangle$ in Eqs.~(\ref{eq:up+}) and (\ref{eq:up-}), one can check that ${\cal PT}|u^\prime_\pm\rangle = |u^\prime_\mp\rangle$ and ${\cal H}|u^\prime_\pm\rangle = E^\prime_\pm |u^\prime_\pm\rangle$, indicating that the energy eigenstates $|u^\prime_\pm\rangle$ are \emph{not} eigenstates of the ${\cal PT}$ operator. This is the reason why this scenario is called the PT-broken phase. However, this is not in contradiction with the fact that $[{\cal PT}, {\cal H}] = {\bf 0}$. Since ${\cal PT}$ is an anti-linear operator and $E^\prime_+ = E^{\prime *}_-$, one should note that ${\cal PT}E^\prime_\pm |u^\prime_\pm \rangle = E^\prime_\mp {\cal PT}|u^\prime_\pm \rangle$. More explicitly, we have
\begin{equation}
{\cal PT} {\cal H} |u^\prime_\pm\rangle = {\cal PT} E^\prime_\pm |u^\prime_\pm \rangle = E^\prime_\mp |u^\prime_\mp \rangle \; , \quad
{\cal H} {\cal PT} |u^\prime_\pm\rangle = {\cal H} |u^\prime_\mp \rangle = E^\prime_\mp |u^\prime_\mp \rangle \; ,
\end{equation}
implying $[{\cal PT}, {\cal H}] = {\bf 0}$. This is quite different from the PT-symmetric phase, in which the two eigenvalues $E^{}_\pm$ are real.

Now, we apply the bi-orthogonal formalism to the non-Hermitian Hamiltonian system in the PT-broken phase. As before~\cite{Ohlsson:2019noy}, we have to find out the eigenvectors of ${\cal H}^\dagger$, namely,
\begin{equation}
{\cal H}^\dagger |v^\prime_\pm\rangle = E^\prime_\pm |v^\prime_\pm \rangle \; . \label{eq:Hdagger_bp}
\end{equation}
The identity ${\cal P}{\cal H}^\dagger {\cal P}^{-1} = {\cal H}$ is still applicable, so we multiply Eq.~(\ref{eq:Hdagger_bp}) on both sides from the left by the ${\cal P}$ operator and then obtain
\begin{equation}
\left({\cal P} {\cal H}^\dagger {\cal P}^{-1} \right) {\cal P} |v^\prime_\pm \rangle = E^\prime_\pm {\cal P}|v^\prime_\pm \rangle \; ,
\label{eq:leftright}
\end{equation}
indicating ${\cal P}|v^\prime_\pm\rangle \propto |u^\prime_\pm \rangle$. Identifying $|v^\prime_\pm \rangle = {\cal P}|u^\prime_\pm\rangle$, we can immediately compute the metric operator $\eta^\prime$, i.e.,
\begin{equation}
\eta^\prime \equiv \sum_{s = \pm} |v^\prime_{+s}\rangle \langle v^\prime_{-s}| = {\cal P} \left(|u^\prime_+ \rangle \langle u^\prime_-| + |u^\prime_-\rangle \langle u^\prime_+|\right) {\cal P} = \left(\begin{matrix} 0 & 1 \cr 1 & 0 \end{matrix}\right)
\end{equation}
and its inverse
\begin{equation}
\eta^{\prime -1} = \sum_{s = \pm} |u^\prime_{+s}\rangle \langle u^\prime_{-s}| = \left(\begin{matrix} 0 & 1 \cr 1 & 0 \end{matrix}\right) \; .
\end{equation}
Note that $\eta^\prime = \eta^{\prime -1} = {\cal P}$ with ${\rm det}~\eta^\prime = -1 < 0$, so it is not positive-definite. In this case, it is impossible to find a Hermitian matrix to convert the non-Hermitian Hamiltonian into a Hermitian one via a similarity transformation. Furthermore, the ${\cal C}$ operator is given by ${\cal C} = {\cal P}^{-1}\eta^\prime = \eta^{\prime -1} {\cal P} = \mathbb{1}_2$, which turns out to be the trivial $2 \times 2$ identity matrix.

Similar to the PT-symmetric phase, we can define the $\eta^\prime$-inner product as well as the ${\cal CPT}$-inner product as follows~\footnote{Strictly speaking, since the operators $\eta^\prime$ and ${\cal CPT}$ (or equivalently ${\cal PT}$) are not positive-definite in the PT-broken phase, it is inappropriate to define the inner product and the norm of state vectors by using these operators. However, since the notion of the ${\cal PT}$-inner product has been widely used in the literature, we follow this definition and implement it to calculate the transition amplitudes and probabilities in Sec.~\ref{sec: broken}. From Eqs.~(\ref{eq:Hdagger_bp}) and (\ref{eq:leftright}), one can observe that the left- and right-eigenvectors are related via $|v^\prime_\pm\rangle = {\cal P}|u^\prime_\pm\rangle$, which means that the usage of the ${\cal PT}$-inner product is equivalent to that of left- and right-eigenvectors in our calculations.}
\begin{itemize}
\item {\it The $\eta^\prime$-inner product} for any two state vectors $|\psi\rangle$ and $|\chi\rangle$ reads
    \begin{equation}
    \langle \psi|\chi \rangle^{}_{\eta^\prime} \equiv \langle \psi| \eta^\prime |\chi \rangle = |\psi\rangle^\dagger \cdot \eta^\prime \cdot |\chi \rangle \; .
    \end{equation}

\item {\it The ${\cal CPT}$-inner product} for any two state vectors $|\psi\rangle$ and $|\chi\rangle$ reads
        \begin{equation}
    \langle \psi|\chi \rangle^{}_{\cal CPT} \equiv \left({\cal CPT}|\psi\rangle\right)^{\rm T} \cdot |\chi \rangle = |\psi\rangle^\dagger \cdot {\cal PC} \cdot |\chi \rangle = \langle \psi|\chi \rangle^{}_{\eta^\prime} \; ,
    \end{equation}
    where ${\cal PC} = \eta^\prime = {\cal P}$ has been used in the last step.
\end{itemize}
Therefore, the two inner products are equivalent and we will not distinguish between them. In addition, since the ${\cal C}$ operator is trivial, these two inner products are also identical with the ${\cal PT}$-inner product. However, as we have mentioned, the metric operator $\eta^\prime = {\cal P}$ is no longer positive-definite, and thus, the norm $\langle \psi|\psi\rangle^{}_{\eta^\prime}$ cannot be guaranteed to be positive. In fact, for the energy eigenstates $|u^\prime_+\rangle$ and $|u^\prime_-\rangle$, we have $\langle u^\prime_\pm | u^\prime_\pm \rangle^{}_{\cal PT} = \langle u^\prime_\pm|u^\prime_\pm\rangle^{}_{\eta^\prime} = 0$ and $\langle u^\prime_\pm| u^\prime_\mp\rangle^{}_{\cal PT} = \langle u^\prime_\pm| u^\prime_\mp \rangle^{}_{\eta^\prime} =  + 1$. The identity $\eta^\prime {\cal H} \eta^{\prime -1} = {\cal H}^\dagger$, which now coincides with ${\cal P} {\cal H} {\cal P}^{-1} = {\cal H}^\dagger$, indeed leads to a unitary time evolution of the energy eigenstates.

\subsection{Exceptional Point}
\label{sub:EP}

In general, let us investigate the exceptional point (EP) of a non-Hermitian Hamiltonian of an open quantum system on the following form (see, e.g., Ref.~\cite{Ozdemir:2019})
\begin{equation}
{\cal H}_{\rm NH} =  \left(\begin{matrix} \omega_1 - {\rm i} \gamma_1 & \kappa \cr \kappa & \omega_2 - {\rm i} \gamma_2 \end{matrix}\right) \; , \label{eq:H_NH}
\end{equation}
where $\omega_1$ and $\omega_2$ are the energies of the states of the system, $\gamma_1$ and $\gamma_2$ are real loss ($\gamma_i > 0$) or gain ($\gamma_i < 0$) parameters (i.e., the widths), and $\kappa$ is the (real) coupling strength between the states of the system, which has the two complex eigenvalues
\begin{equation}
\lambda_\pm = \frac{1}{2} \left[ \omega_1 + \omega_2 - {\rm i} \left( \gamma_1 + \gamma_2 \right) \right] \pm \sqrt{Q}
\end{equation}
with $Q$ being the quadratic expression
\begin{equation}
Q = \kappa^2 + \frac{1}{4} \left[ \left( \omega_1 - \omega_2 \right)^2 - \left( \gamma_1 - \gamma_2 \right)^2 \right] - \frac{{\rm i}}{2} \left( \omega_1 - \omega_2 \right) \left( \gamma_1 - \gamma_2 \right) \; .
\end{equation}
In order for the non-Hermitian Hamiltonian in Eq.~(\ref{eq:H_NH}) to have EPs, the condition $Q = 0$ must be fulfilled and this condition will determine the EPs that give rise to one degenerate eigenvalue.

Now, there are several specific cases that are interesting to study of the Hamiltonian in Eq.~(\ref{eq:H_NH}), which are (i) the case $\omega_1 = \omega_2 = \omega_0$, (ii) a non-Hermitian Hamiltonian consisting of a so-called {\it passive} PT-symmetric Hamiltonian and a Hamiltonian for a {\it lossy} uncoupled system \cite{Joglekar:2018,Ozdemir:2019} with $\omega_1 = \omega_2 = \omega$, $\gamma_1 = \gamma + \chi$, and $\gamma_2 = - \gamma + \chi$, where $\chi$ is a global decay parameter, (iii) a PT-symmetric Hamiltonian with $\omega_1 = \omega_2 = \omega$, $\gamma_1 = - \gamma_2 = \gamma$, and (iv) the case $\omega_1 = \omega_2 = \omega'$, $\gamma_1 = \gamma'$, and $\gamma_2 = 0$. In the following, we describe the four different cases in detail. First, in case~(i), the two complex eigenvalues and quadratic expression reduce to
\begin{equation}
\lambda_\pm = \omega_0 - \frac{{\rm i}}{2} \left( \gamma_1 + \gamma_2 \right) \pm \sqrt{Q}, \qquad Q = \kappa^2 - \frac{1}{4} \left( \gamma_1 - \gamma_2 \right)^2 \; ,
\end{equation}
and therefore, the condition to have EPs is given by $\kappa^2 = (\gamma_1 - \gamma_2)^2/4$. In Ref.~\cite{Roccati:2021}, quantum corrections to non-Hermitian systems at the EPs have been studied. Second, for the non-Hermitian Hamiltonian including the passive PT-symmetric Hamiltonian [i.e., case~(ii)], we obtain
\begin{equation}
\lambda_\pm = \omega - {\rm i} \chi \pm \sqrt{Q}, \qquad Q = \kappa^2 - \gamma^2,
\end{equation}
so the condition for EPs is simply $\kappa^2 = \gamma^2$, which means that variation of a single parameter is sufficient to reach the EPs. In Ref.~\cite{Joglekar:2018}, it has been shown that passive PT transitions in dissipative photonic systems are however not dependent on the existence of EPs. Note that the condition for the EPs is independent of $\chi$ (i.e., the global decay or damping of the system). At the EPs, the Hamiltonian in Eq.~(\ref{eq:H_NH}) will only have damping (described by $\chi$) and be independent of both $\gamma$ and $\kappa$. Thus, at the EPs, the Hamiltonian will effectively be
\begin{equation}
{\cal H}_{\rm NH}^{\rm eff.} = \left(\begin{matrix} \omega - {\rm i} \chi & 0 \cr 0 & \omega - {\rm i} \chi \end{matrix}\right) \; .
\end{equation}
Case~(ii) has been reviewed in Ref.~\cite{Ozdemir:2019} for PT-symmetric systems and their EPs in photonics (see also the review in Ref.~\cite{Feng:2017}), and will also be investigated in this work in connection to the neutral-meson system.\footnote{Note that the dynamics of a loss-loss system is basically equivalent to that of the combination of a loss-gain system and a system with global exponential decay, see, e.g., Ref.~\cite{Ozdemir:2019}.} Third, for the PT-symmetric Hamiltonian [i.e., case~(iii)], the parameter $\chi = 0$ in case~(ii), so we find that
\begin{equation}
\lambda_\pm = \omega \pm \sqrt{Q}, \qquad Q = \kappa^2 - \gamma^2,
\end{equation}
which means that the condition for EPs is again $\kappa^2 = \gamma^2$. In Ref.~\cite{Eleuch:2014}, the EPs in open and PT-symmetric systems have been studied. For example, such a two-state quantum system could exhibit both loss and gain described by the parameters $\gamma_1 = \gamma$ and $\gamma_2 = - \gamma$, respectively, and coupling of the two states via the lattice with the environment. Furthermore, in Ref.~\cite{Wang:2019}, the application of the PT dimer in photonics and phononics, which exhibits EPs, has been discussed. Fourth, in case~(iv), we have
\begin{equation}
\lambda_\pm = \omega' - \frac{{\rm i}}{2} \gamma' \pm \sqrt{Q}, \qquad Q = \kappa^2 - \frac{1}{4} {\gamma'}^2,
\end{equation}
and the EPs are given by the condition $\kappa^2 = {\gamma'}^2/4$. This case is similar to case~(iii), but with $\gamma_2 = 0$, i.e., vanishing gain, which has also been studied in Ref.~\cite{Eleuch:2014}. In addition, the dynamics and EPs of this type of Hamiltonian has been discussed in Ref.~\cite{Wang:2020}. 
Note that although case~(iv) is similar to case~(iii), it exhibits a rather different behavior for the degenerate eigenvalue at the EPs, which is more similar to case~(ii) with the important difference that the imaginary part of the eigenvalue is related to the condition for the EPs.

Finally, for the most general PT-symmetric non-Hermitian Hamiltonian in Eq.~(\ref{eq:H}) with $\phi = 0$, which is the same as case~(ii) with $\chi = 0$ or case~(iii), let us give a brief discussion about the EP at $\rho^2\sin^2\varphi = \sigma^2$. The EP can be identified as either the limiting case of $\alpha \to 0$ in the PT-symmetric phase or that of $\alpha^\prime \to 0$ in the PT-broken phase. In either limit, the energy eigenvalues become degenerate $E^{}_\pm$ (or $E^\prime_\pm$) $\to E^{}_0 = \rho \cos\varphi$. Moreover, for the eigenvectors $|u^{}_\pm\rangle$ in Eq.~(\ref{eq:u+u-2}) and $|u^\prime_\pm\rangle$ in Eqs.~(\ref{eq:up+}) and (\ref{eq:up-}), the normalization constants $N^{}_\pm \propto 1/\sqrt{2\sin\alpha}$ and $N^\prime_\pm \propto 1/\sqrt{2\sinh\alpha^\prime}$ are divergent in the respective limits of $\alpha \to 0$ and $\alpha^\prime \to 0$. However, this is an artificial divergence, since $N^{}_\pm$ (or $N^\prime_\pm$) in the limit of $\alpha \to 0$ (or $\alpha^\prime \to 0$) cannot be determined from the ${\cal PT}$-inner products of the relevant eigenvectors. The proper normalization can be taken as $\langle u^{}_0|u^{}_0\rangle = 1$, with $\langle u^{}_0| \equiv |u^{}_0\rangle^\dagger$, so we have
\begin{equation}
|u^{}_\pm\rangle ~ ({\rm or}~|u^\prime_\pm\rangle) \to |u^{}_0\rangle = \frac{1}{\sqrt{2}} \left(\begin{matrix} e^{+{\rm i}\pi/4} \cr e^{-{\rm i}\pi/4} \end{matrix}\right) \; ,
\end{equation}
corresponding to the degenerate eigenvalue $E^{}_0$ at the EP. The rich physics at the EPs and their practical applications have been briefly summarized in Refs.~\cite{Ozdemir:2019, Berry, Heiss:2012dx, Wiersig1, Wiersig2, Sensors}.

Since the time evolution of $|u^{}_0\rangle$ is governed by the Schr\"{o}dinger equation, we have $|u^{}_0(t)\rangle = e^{-{\rm i}E^{}_0 t}|u^{}_0\rangle$, implying that only an overall phase factor will develop and no transitions between any two quantum states are expected. This is also true for the flavor eigenstates $|u^{}_a\rangle = (1, 0)^{\rm T}$ and $|u^{}_b\rangle = (0, 1)^{\rm T}$, which are linear superpositions of the energy eigenstates.

\section{Transitions in the PT-Broken Phase}\label{sec: broken}

\subsection{PT-Inner Product}
\label{sub: PTip}

Since the transition amplitudes and probabilities between two flavor eigenstates in the PT-symmetric phase have been examined in detail in Ref.~\cite{Ohlsson:2019noy}, we now consider the transitions between two flavor eigenstates in the PT-broken phase in this section. In this scenario, the Schr\"{o}dinger equation for the time evolution of the energy eigenstates is
\begin{equation}
{\rm i}\frac{{\rm d}}{{\rm d}t} |u^\prime_\pm(t)\rangle = {\cal H} |u^\prime_\pm(t)\rangle = E^\prime_\pm |u^\prime_\pm\rangle \; ,
\end{equation}
and thus, we have
\begin{eqnarray}
|u^\prime_+(t)\rangle &=& e^{-{\rm i}E^\prime_+ t} |u^\prime_+(0)\rangle = \frac{e^{-{\rm i}\omega t + \gamma t}}{\sqrt{2\sinh \alpha^\prime}}  \left(\begin{matrix} e^{+\alpha^\prime/2} \cr -{\rm i} e^{-\alpha^\prime/2} \end{matrix}\right) \; , \\
|u^\prime_-(t)\rangle &=& e^{-{\rm i}E^\prime_- t} |u^\prime_-(0)\rangle = \frac{e^{-{\rm i}\omega t - \gamma t}}{\sqrt{2\sinh \alpha^\prime}}  \left(\begin{matrix} +{\rm i} e^{-\alpha^\prime/2} \cr  e^{+\alpha^\prime/2} \end{matrix}\right) \; ,
\end{eqnarray}
where the auxiliary parameters $\omega \equiv \rho \cos\varphi$ and $\gamma \equiv \sqrt{\rho^2\sin^2\varphi - \sigma^2}$ have been defined. For reference, we list below the correspondences between the three new parameters $\{\omega, \gamma, \alpha^\prime\}$ and the three original ones $\{\rho, \sigma, \varphi\}$, where $\phi = 0$ has been assumed as in Sec.~\ref{sec: general},
\begin{equation}
\omega = \rho \cos\varphi \; , \quad \gamma = \sqrt{\rho^2\sin^2\varphi - \sigma^2} \; , \quad \alpha^\prime = {\rm arcosh} \left(\frac{\rho\sin\varphi}{\sigma}\right)
\end{equation}
or
\begin{equation}
\rho = \sqrt{\omega^2 + \gamma^2 \coth^2\alpha^\prime} \; , \quad \sigma = \frac{\gamma}{\sinh \alpha^\prime} \; , \quad \varphi = \arccos \left(\frac{\omega}{\sqrt{\omega^2 + \gamma^2 \coth^2 \alpha^\prime}}\right) \; . \label{eq:rsf}
\end{equation}
In the following, we adopt the new set of parameters $\{\omega, \gamma, \alpha^\prime\}$, which can be converted back to the original one by using Eq.~(\ref{eq:rsf}).

To demonstrate the unitary time evolution, we calculate the norms of the time-evolved energy eigenstates to find
\begin{eqnarray}
\langle u^\prime_+(t) | u^\prime_+(t)\rangle^{}_{\cal PT} = |u^\prime_+(t)\rangle^\dagger \cdot {\cal P} \cdot |u^\prime_+(t)\rangle = 0 \; , \\
\langle u^\prime_-(t) | u^\prime_-(t)\rangle^{}_{\cal PT} = |u^\prime_-(t)\rangle^\dagger \cdot {\cal P} \cdot |u^\prime_-(t)\rangle = 0 \; .
\end{eqnarray}
Similarly, one can also verify that $\langle u^\prime_\pm(t)|u^\prime_\mp(t)\rangle^{}_{\cal PT} = +1$, which is time-independent as it should be.

Next, we introduce the flavor eigenstates in which basis the explicit form of the non-Hermitian Hamiltonian is specified. Recall the diagonalization of the Hamiltonian, i.e.,
\begin{equation}
A^\prime {\cal H} A^{\prime -1} = \widehat{\cal H} \equiv \left(\begin{matrix} E^\prime_+ & 0 \cr 0 & E^\prime_- \end{matrix}\right) \quad \Longrightarrow \quad ({\cal H} |w^{}_+\rangle, {\cal H} |w^{}_-\rangle) = (|w^{}_+\rangle E^\prime_+, |w^{}_-\rangle E^\prime_-) \; ,
\end{equation}
where we have written $A^{\prime -1} = (|w^{}_+\rangle, |w^{}_-\rangle)$ with $|w^{}_\pm\rangle$ being two column vectors. Obviously, we can identify $|w^{}_\pm\rangle$ with $|u^\prime_\pm\rangle$ in Eqs.~(\ref{eq:up+}) and (\ref{eq:up-}), since ${\cal H}|u^\prime_\pm\rangle = E^\prime_\pm |u^\prime_\pm\rangle$. Hence, it is easy to derive
\begin{equation}
A^{\prime -1} = \frac{1}{\sqrt{2\sinh \alpha^\prime}}  \left(\begin{matrix} e^{+\alpha^\prime/2} & +{\rm i} e^{-\alpha^\prime/2} \cr -{\rm i} e^{-\alpha^\prime/2} & e^{+\alpha^\prime/2} \end{matrix}\right) \; , \quad A^\prime = \frac{1}{\sqrt{2\sinh \alpha^\prime}}  \left(\begin{matrix} e^{+\alpha^\prime/2} & -{\rm i} e^{-\alpha^\prime/2} \cr +{\rm i} e^{-\alpha^\prime/2} & e^{+\alpha^\prime/2} \end{matrix}\right) \; ,
\end{equation}
where one can note that $A^{\prime -1} = A^{\prime \rm T}$ and $A^{\prime \dagger} = A^\prime$. Furthermore, it is straightforward to verify that the flavor eigenstates are given by
\begin{eqnarray}
|u^\prime_a\rangle &=& \left(A^{\prime -1}\right)^{}_{a+} |u^\prime_+\rangle + \left(A^{\prime -1}\right)^{}_{a-} |u^\prime_-\rangle = \left(\begin{matrix} 1 \cr 0 \end{matrix}\right) \; , \label{eq:upa}\\
|u^\prime_b\rangle &=& \left(A^{\prime -1}\right)^{}_{b+} |u^\prime_+\rangle + \left(A^{\prime -1}\right)^{}_{b-} |u^\prime_-\rangle = \left(\begin{matrix} 0 \cr 1 \end{matrix}\right) \; , \label{eq:upb}
\end{eqnarray}
which resemble the forms in the PT-symmetric phase. This should be the case as the explicit form of the Hamiltonian in Eq.~(\ref{eq:H}) remains the same in the PT-broken phase. In Eqs.~(\ref{eq:upa}) and (\ref{eq:upb}), $\left(A^{\prime -1}\right)^{}_{\beta s}$ for $\beta = a, b$ and $s = +, -$ denote the matrix elements of $A^{\prime -1}$. One can also prove that the norms $\langle u^\prime_a(t)|u^\prime_a(t)\rangle^{}_{\cal PT} = \langle u^\prime_b(t)|u^\prime_b(t)\rangle^{}_{\cal PT} = 0$ and $\langle u^\prime_a(t)|u^\prime_b(t)\rangle^{}_{\cal PT} = \langle u^\prime_b(t)|u^\prime_a(t)\rangle^{}_{\cal PT} = +1$ are time-independent.

Then, we proceed to compute the amplitudes and probabilities for the transitions between two flavor eigenstates. After some calculations, the transition amplitudes are found to be
\begin{eqnarray}
&~& {\cal A}^\prime_{aa} \equiv \langle u^\prime_a|u^\prime_a(t)\rangle^{}_{\cal PT} = -{\rm i} e^{-{\rm i}\omega t} \frac{\sinh(\gamma t)}{\sinh \alpha^\prime} \; , \label{eq:Apaa}\\
&~& {\cal A}^\prime_{ab} \equiv \langle u^\prime_b|u^\prime_a(t)\rangle^{}_{\cal PT} = e^{-{\rm i}\omega t} \frac{\sinh(\alpha^\prime + \gamma t)}{\sinh \alpha^\prime} \; , \label{eq:Apab}\\
&~& {\cal A}^\prime_{ba} \equiv \langle u^\prime_a|u^\prime_b(t)\rangle^{}_{\cal PT} = e^{-{\rm i}\omega t} \frac{\sinh(\alpha^\prime - \gamma t)}{\sinh \alpha^\prime} \; , \label{eq:Apba}\\
&~& {\cal A}^\prime_{bb} \equiv \langle u^\prime_b|u^\prime_b(t)\rangle^{}_{\cal PT} = -{\rm i} e^{-{\rm i}\omega t} \frac{\sinh(\gamma t)}{\sinh \alpha^\prime} \; , \label{eq:Apbb}
\end{eqnarray}
while the corresponding transition probabilities are defined as ${\cal P}^\prime_{\alpha \beta} \equiv |{\cal A}^\prime_{\alpha \beta}|^2$ (for $\alpha, \beta$ running over $a, b$) and explicitly calculated as
\begin{eqnarray}
&~& {\cal P}^\prime_{aa} = \sinh^2(\gamma t)/\sinh^2 \alpha^\prime \; , \label{eq:Ppaa}\\
&~& {\cal P}^\prime_{ab} = \sinh^2(\alpha^\prime + \gamma t)/\sinh^2 \alpha^\prime \; , \label{eq:Ppab}\\
&~& {\cal P}^\prime_{ba} = \sinh^2(\alpha^\prime - \gamma t)/\sinh^2 \alpha^\prime \; , \label{eq:Ppba}\\
&~& {\cal P}^\prime_{bb} = \sinh^2(\gamma t)/\sinh^2 \alpha^\prime \; . \label{eq:Ppbb}
\end{eqnarray}
One can observe non-conservation of the total probability, i.e., ${\cal P}^\prime_{aa} + {\cal P}^\prime_{ab} \neq 1$ or ${\cal P}^\prime_{ba} + {\cal P}^\prime_{bb} \neq 1$. Moreover, all probabilities in Eqs.~(\ref{eq:Ppaa})--(\ref{eq:Ppbb}) go to infinity for $t \to +\infty$, rendering them to be physically meaningless. However, the metric operator $\eta^\prime = {\cal P}$ is not positive-definite, so we should not expect the sum of transition probabilities to be conserved. One may instead compute the differences between the probabilities, i.e.,
\begin{eqnarray}
{\cal P}^\prime_{aa} - {\cal P}^\prime_{ab} &=& - \left[\sinh^2(\alpha^\prime + \gamma t) - \sinh^2 (\gamma t)\right]/\sinh^2 \alpha^\prime = - \sinh(\alpha^\prime + 2\gamma t)/\sinh \alpha^\prime \; , \\
{\cal P}^\prime_{ba} - {\cal P}^\prime_{bb} &=& + \left[\sinh^2(\alpha^\prime - \gamma t) - \sinh^2 (\gamma t)\right]/\sinh^2 \alpha^\prime = + \sinh(\alpha^\prime - 2\gamma t)/\sinh \alpha^\prime \; , \quad
\end{eqnarray}
which are unfortunately time-dependent. As a remedy for this problem, following the same strategy as in Ref.~\cite{Ohlsson:2019noy}, we construct the ${\cal CPT}$ flavor eigenstates $|\tilde{u}^\prime_a\rangle$ and $|\tilde{u}^\prime_b\rangle$ as follows
\begin{eqnarray}
|\tilde{u}^\prime_a\rangle &=& \frac{1}{\sqrt{2}} \left(|u^\prime_a\rangle + |u^\prime_b\rangle \right) = \frac{1}{\sqrt{2}} \left(\begin{matrix} +1 \cr +1 \end{matrix}\right)\; , \\
|\tilde{u}^\prime_b\rangle &=& \frac{1}{\sqrt{2}} \left(|u^\prime_a\rangle - |u^\prime_b\rangle \right) = \frac{1}{\sqrt{2}} \left(\begin{matrix} +1 \cr -1 \end{matrix}\right)\; ,
\end{eqnarray}
where the expressions for the two original flavor eigenstates $|u^\prime_a\rangle$ and $|u^\prime_b\rangle$ in Eqs.~(\ref{eq:upa}) and (\ref{eq:upb}) have been used. Since the ${\cal C}$ operator is trivial in the PT-broken phase, we can easily prove that ${\cal CPT}|\tilde{u}^\prime_a\rangle = {\cal PT}|\tilde{u}^\prime_a\rangle = + |\tilde{u}^\prime_a\rangle$ and ${\cal CPT}|\tilde{u}^\prime_b\rangle = {\cal PT}|\tilde{u}^\prime_b\rangle = - |\tilde{u}^\prime_b\rangle$. Therefore, the newly-constructed flavor eigenstates are eigenstates of both the ${\cal CPT}$ and ${\cal PT}$ operators. With these ${\cal CPT}$ flavor eigenstates, we repeat the calculations of the transition amplitudes and probabilities, and then obtain the amplitudes $\tilde{\cal A}^\prime_{\alpha \beta} \equiv \langle \tilde{u}^\prime_\beta|u^\prime_\alpha(t)\rangle$ as
\begin{eqnarray}
&~& \tilde{\cal A}^\prime_{aa} \equiv \langle \tilde{u}^\prime_a|u^\prime_a(t)\rangle^{}_{\cal PT} = \frac{1}{\sqrt{2}} \left({\cal A}^\prime_{aa} + {\cal A}^\prime_{ab}\right) \; , \\
&~& \tilde{\cal A}^\prime_{ab} \equiv \langle \tilde{u}^\prime_b|u^\prime_a(t)\rangle^{}_{\cal PT} = \frac{1}{\sqrt{2}} \left({\cal A}^\prime_{aa} - {\cal A}^\prime_{ab}\right) \; , \\
&~& \tilde{\cal A}^\prime_{ba} \equiv \langle \tilde{u}^\prime_a|u^\prime_b(t)\rangle^{}_{\cal PT} = \frac{1}{\sqrt{2}} \left({\cal A}^\prime_{ba} + {\cal A}^\prime_{bb}\right) \; , \\
&~& \tilde{\cal A}^\prime_{bb} \equiv \langle \tilde{u}^\prime_b|u^\prime_b(t)\rangle^{}_{\cal PT} = \frac{1}{\sqrt{2}} \left({\cal A}^\prime_{ba} - {\cal A}^\prime_{bb}\right) \; ,
\end{eqnarray}
and the probabilities $\tilde{\cal P}^\prime_{\alpha \beta} \equiv |\tilde{\cal A}^\prime_{\alpha \beta}|^2$ as
\begin{eqnarray}
&~& \tilde{\cal P}^\prime_{aa} = \tilde{\cal P}^\prime_{ab} = \frac{1}{2} \left({\cal P}^\prime_{aa} + {\cal P}^\prime_{ab}\right) = \frac{1}{2\sinh^2\alpha^\prime} \left[\sinh^2 (\gamma t) + \sinh^2 (\alpha^\prime + \gamma t)\right] \; , \label{eq:Ptpaa}\\
&~& \tilde{\cal P}^\prime_{ba} = \tilde{\cal P}^\prime_{bb} = \frac{1}{2} \left({\cal P}^\prime_{ba} + {\cal P}^\prime_{bb}\right) = \frac{1}{2\sinh^2\alpha^\prime} \left[\sinh^2 (\gamma t) + \sinh^2 (\alpha^\prime - \gamma t)\right] \; . \label{eq:Ptpba}
\end{eqnarray}
Although these probabilities still become infinite in the limit of $t \to +\infty$, one can check that $\tilde{\cal P}^\prime_{aa} - \tilde{\cal P}^\prime_{ab} = 0$ and $\tilde{\cal P}^\prime_{ba} - \tilde{\cal P}^\prime_{bb} = 0$, where the time dependence is completely canceled out. Thus, the introduction of the ${\cal CPT}$ flavor eigenstates basically renormalizes the individual transition probabilities in Eqs.~(\ref{eq:Ptpaa}) and (\ref{eq:Ptpba}) for the flavor eigenstates and reduces the effect of $\eta'$, as discussed in Ref.~\cite{Ohlsson:2019noy}.

It is interesting to note that there is no interference between the two amplitudes ${\cal A}^\prime_{aa}$ and ${\cal A}^\prime_{ab}$ when squaring the modified amplitudes $\tilde{\cal A}^\prime_{aa}$ and $\tilde{\cal A}^\prime_{ab}$ to calculate $\tilde{\cal P}^\prime_{aa}$ and $\tilde{\cal P}^\prime_{ab}$, leading to a simple average of the probabilities in Eq.~(\ref{eq:Ptpaa}). The main reason can be traced back to the amplitudes in Eqs.~(\ref{eq:Apaa}) and (\ref{eq:Apab}), where ${\cal A}^\prime_{aa}$ is purely imaginary, whereas ${\cal A}^\prime_{ab}$ is real up to the same phase factor $e^{-{\rm i}\omega t}$. Similar observations can be made for $\tilde{\cal P}^\prime_{ba}$ and $\tilde{\cal P}^\prime_{bb}$ in Eq.~(\ref{eq:Ptpba}). Therefore, it seems more reasonable to define the ${\cal CPT}$ flavor eigenstates as the final states in the sense of the time-independence of the probability differences.

\subsection{Connection between the PT Symmetry and the Neutral-Meson System}

The non-Hermitian Hamiltonian with complex eigenvalues has been known in particle physics for a long time. As a concrete example, the mixing and oscillation of the neutral-meson system $\{|P^0\rangle, |\overline{P}^0\rangle\}$, such as $K^0$-$\overline{K}^0$, $D^0$-$\overline{D}^0$, and $B^0$-$\overline{B}^0$, can be described by an effective non-Hermitian Hamiltonian~\cite{Lee:1957qq, Lee:1965hi, Bigi:2000yz}
\begin{equation}
{\sf H} = {\sf M} - \frac{\rm i}{2} {\sf \Gamma} \equiv \left(\begin{matrix} M^{}_{11} & M^{}_{12} \cr M^*_{12} & M^{}_{22}\end{matrix}\right) - \frac{\rm i}{2} \left(\begin{matrix} \Gamma^{}_{11} & \Gamma^{}_{12} \cr \Gamma^*_{12} & \Gamma^{}_{22}\end{matrix}\right) \; , \label{eq:HMG}
\end{equation}
where both ${\sf M}$ and ${\sf \Gamma}$ are $2\times 2$ Hermitian matrices. In order to make a distinction between the neutral-meson system and the PT-broken phase under consideration, we have set all $2\times 2$ matrices in the former case in a sans-serif typeface. Without imposing either CPT or CP invariance,\footnote{Note that the C, P, and T transformations, as well as their combinations CPT and CP, for the neutral-meson system should be understood in the same way as in particle physics or relativistic quantum field theories in general.} the time-evolved neutral-meson states can be written as~\cite{Tanabashi:2018oca}
\begin{eqnarray}
|P^0(t)\rangle &=& \left[g^{}_+(t) + z g^{}_-(t)\right]|P^0\rangle - \frac{q}{p} \sqrt{1 - z^2} g^{}_-(t) |\overline{P}^0\rangle \; , \\
|\overline{P}^0(t)\rangle &=& \left[g^{}_+(t) - z g^{}_-(t)\right]|\overline{P}^0\rangle - \frac{p}{q} \sqrt{1 - z^2} g^{}_-(t) |P^0\rangle \; ,
\end{eqnarray}
where $z = 0$ corresponds to the case of either CPT or CP invariance and the relevant time-evolution functions are given by
\begin{equation}
g^{}_\pm(t) \equiv \frac{1}{2} \left[\exp\left(-{\rm i}M^{}_2 t - \frac{1}{2} \Gamma^{}_2 t\right) \pm \exp\left(-{\rm i}M^{}_1 t - \frac{1}{2} \Gamma^{}_1 t\right) \right] \; .
\end{equation}
Note that for $i = 1, 2$, $M^{}_i$ stand for the masses of the energy eigenstates $|P^{}_i\rangle$, while $\Gamma^{}_i$ for the corresponding total decay widths. The masses and decay widths, which all should be positive (i.e., $M_i > 0$ and $\Gamma_i > 0$) for the neutral-meson system, are related to the matrix elements of the effective Hamiltonian with the eigenvalues $\{E^{}_1, E^{}_2\}$ via
\begin{eqnarray}
E^{}_1 &\equiv& M^{}_1 - \frac{\rm i}{2} \Gamma^{}_1 = M^{}_{11} - \frac{\rm i}{2} \Gamma^{}_{11} + pq \left[\kappa + \sqrt{1 + \kappa^2}\right] \; , \label{eq:E1}\\
E^{}_2 &\equiv& M^{}_2 - \frac{\rm i}{2} \Gamma^{}_2 = M^{}_{22} - \frac{\rm i}{2} \Gamma^{}_{22} - pq \left[\kappa + \sqrt{1 + \kappa^2}\right] \; , \label{eq:E2}
\end{eqnarray}
where $\kappa \equiv \left[(M^{}_{22} - {\rm i}\Gamma^{}_{22}/2) - (M^{}_{11} - {\rm i}\Gamma^{}_{11}/2) \right]/(2pq)$ and
\begin{equation}
p^2 \equiv M^{}_{12} - \frac{\rm i}{2}\Gamma^{}_{12} \; , \quad q^2 \equiv M^*_{12} - \frac{\rm i}{2}\Gamma^*_{12} \; .
\end{equation}
The complex parameter $z$ can be expressed as follows
\begin{equation}
z \equiv \frac{\kappa}{\sqrt{1 + \kappa^2}} = \frac{\displaystyle \delta m - \frac{\rm i}{2}\delta \Gamma}{\displaystyle \Delta m - \frac{\rm i}{2}\Delta \Gamma} \label{eq:z}
\end{equation}
with $\delta m \equiv M^{}_{11} - M^{}_{22}$, $\Delta m \equiv M^{}_2 - M^{}_1$, $\delta \Gamma \equiv \Gamma^{}_{11} - \Gamma^{}_{22}$, and $\Delta \Gamma \equiv \Gamma^{}_2 - \Gamma^{}_1$. Now, it is evident that $z = 0$ corresponds to $M^{}_{11} = M^{}_{22}$ and $\Gamma^{}_{11} = \Gamma^{}_{22}$, as implied by the CPT theorem for local quantum field theories.

It is straightforward to calculate the transition amplitudes for $|P^0\rangle \rightarrow |P^0\rangle$ and $|P^0\rangle \rightarrow |\overline{P}^0\rangle$, namely,
\begin{eqnarray}
{\cal A}^{}_{P^0 P^0}(t) &\equiv& \langle P^0|P^0(t)\rangle = g^{}_+(t) + z g^{}_-(t) \; , \\
{\cal A}^{}_{P^0 \overline{P}^0}(t) &\equiv& \langle \overline{P}^0|P^0(t)\rangle = - \frac{q}{p} \sqrt{1 - z^2}  g^{}_-(t) \; ,
\end{eqnarray}
where $\langle P^0| \equiv |P^0\rangle^\dagger$ and $\langle \overline{P}^0| \equiv |\overline{P}^0\rangle^\dagger$ have been defined. Accordingly, the corresponding transition probabilities turn out to be
\begin{eqnarray}
{\cal P}^{}_{P^0 P^0}(t) \equiv |{\cal A}^{}_{P^0 P^0}(t)|^2 &=& + \frac{1}{4} \left[e^{-\Gamma^{}_1t} + e^{-\Gamma^{}_2t} + 2 e^{-\Gamma t} \cos (\Delta m t) \right] \nonumber \\
&~& + \frac{1}{4} \left[e^{-\Gamma^{}_1t} + e^{-\Gamma^{}_2t} - 2 e^{-\Gamma t} \cos (\Delta m t) \right] |z|^2  \nonumber \\
&~& + \frac{1}{2} \left(e^{-\Gamma^{}_2t} - e^{-\Gamma^{}_1t}\right) \Re(z) + e^{-\Gamma t} \sin (\Delta m t) \, \Im(z) \; , \label{eq:PP0P0}\\
{\cal P}^{}_{P^0 \overline{P}^0}(t) \equiv |{\cal A}^{}_{P^0 \overline{P}^0}(t)|^2 &=& \frac{|q|^2}{4|p|^2} \left[e^{-\Gamma^{}_1t} + e^{-\Gamma^{}_2t} - 2 e^{-\Gamma t} \cos (\Delta m t) \right] \sqrt{1 - 2 \Re(z^2) + |z|^4}\; , \nonumber\\ \label{eq:PP0bP0}
\end{eqnarray}
with $\Gamma \equiv (\Gamma^{}_1 + \Gamma^{}_2)/2$. Since the decay widths $\Gamma^{}_1$ and $\Gamma^{}_2$ are positive, the transition probabilities ${\cal P}^{}_{P^0 P^0}(t)$ and ${\cal P}^{}_{P^0 \overline{P}^0}(t)$ will vanish in the limit of $t \to +\infty$.

As observed in Ref.~\cite{Ozdemir:2019}, a loss-loss system in photonics can be equivalently described as a PT-symmetric loss-gain system with global exponential decay or amplification. Inspired by this observation of passive PT-symmetry in the loss-loss system, we assume that the most general non-Hermitian Hamiltonian in Eq.~(\ref{eq:HMG}) can be decomposed into a PT-symmetric non-Hermitian Hamiltonian in the PT-broken phase and a lossy term, i.e.,
\begin{eqnarray}
\left(\begin{matrix} M^{}_{11} & M^{}_{12} \cr M^*_{12} & M^{}_{22}\end{matrix}\right) - \frac{\rm i}{2} \left(\begin{matrix} \Gamma^{}_{11} & \Gamma^{}_{12} \cr \Gamma^*_{12} & \Gamma^{}_{22}\end{matrix}\right) = \left(\begin{matrix} \rho e^{+{\rm i}\varphi} & \sigma \cr  \sigma & \rho e^{-{\rm i}\varphi} \end{matrix}\right) - \left( \begin{matrix} {\rm i}\chi & 0 \cr 0 &  {\rm i} \chi \end{matrix} \right) \; ,
\label{eq:H_pass}
\end{eqnarray}
where the first term in the right-hand side is simply the PT-symmetric Hamiltonian in Eq.~(\ref{eq:H}) with $\phi = 0$ and the second one with $\chi > 0$ represents a global exponential decay. By identifying both sides of Eq.~(\ref{eq:H_pass}), we obtain the following relations
\begin{eqnarray}
M^{}_{11} - \frac{\rm i}{2} \Gamma^{}_{11} &=& \rho \cos\varphi + {\rm i} \rho \sin\varphi - {\rm i}\chi \; , \\
M^{}_{22} - \frac{\rm i}{2} \Gamma^{}_{22} &=& \rho \cos\varphi - {\rm i} \rho \sin\varphi - {\rm i}\chi \; , \\
M^{}_{12} - \frac{\rm i}{2} \Gamma^{}_{12} &=& \sigma \; , \\
M^*_{12} - \frac{\rm i}{2} \Gamma^*_{12} &=& \sigma \; ,
\end{eqnarray}
implying that $M^{}_{11} = M^{}_{22} = \rho \cos\varphi$, $\Gamma^{}_{11} = - 2(\rho \sin\varphi - \chi) $, $\Gamma^{}_{22} = 2(\rho \sin\varphi + \chi)$, $M^{}_{12} = \sigma$, and $\Gamma^{}_{12} = 0$, together with $p = q = \sqrt{M^{}_{12}} = \sqrt{\sigma}$ and $\kappa = -{\rm i}(\rho\sin\varphi)/\sigma$. Using Eqs.~(\ref{eq:E1}) and (\ref{eq:E2}), we immediately find that
\begin{eqnarray}
&~& M^{}_1 = \rho \cos\varphi \; , \quad \Gamma^{}_1 = 2 \left(\chi - \sqrt{\rho^2 \sin^2\varphi - \sigma^2} \right) \; , \\
&~& M^{}_2 = \rho \cos\varphi \; , \quad \Gamma^{}_2 = 2 \left(\chi + \sqrt{\rho^2 \sin^2\varphi - \sigma^2} \right) \; ,
\end{eqnarray}
where it should be noted that $\rho^2\sin^2\varphi > \sigma^2$ and $(1 - \rho^2\sin^2\varphi/\sigma^2)^{1/2} = +{\rm i} \sqrt{\rho^2 \sin^2\varphi - \sigma^2}/\sigma$ has been utilized. The parameter $z$ in Eq.~(\ref{eq:z}) is determined by $\delta m \equiv M^{}_{11} - M^{}_{22} = 0$, $\Delta m \equiv M^{}_2 - M^{}_1 = 0$, $\delta \Gamma \equiv \Gamma^{}_{11} - \Gamma^{}_{22} = -4\rho \sin\varphi$, and $\Delta \Gamma \equiv \Gamma^{}_2 - \Gamma^{}_1 = 4\sqrt{\rho^2 \sin^2\varphi - \sigma^2}$, namely,
\begin{equation}
z = \frac{\delta \Gamma}{\Delta \Gamma} = -\frac{\rho \sin\varphi}{\sqrt{\rho^2 \sin^2\varphi - \sigma^2}} = - \coth \alpha^\prime \; .
\end{equation}
From the previous discussion, one can recognize that $E^{}_1 = E^\prime_+ = \omega + {\rm i}(\gamma - \chi)$ and $E^{}_2 = E^\prime_- = \omega - {\rm i}(\gamma + \chi)$ with $\omega = \rho \cos\varphi$ and $\gamma = \sqrt{\rho^2 \sin^2 \varphi - \sigma^2}$, and thus, we obtain $M^{}_1 = M^{}_2 = \omega$, $\Gamma^{}_1 = 2(\chi - \gamma)$, and $\Gamma^{}_2 = 2 (\chi + \gamma)$. Since $\Gamma^{}_1$ should be identified with a positive decay width, we have $\Gamma^{}_1 > 0$, or equivalently, $\chi > \gamma$. This condition, if expressed in terms of the averaged decay width $\Gamma = 2\chi$ and the decay-width difference $\Delta \Gamma = 4\gamma$, implies that $\Gamma > \Delta \Gamma/2$, and naturally, it also holds that $\Gamma_2  > 0$, since $\chi > \gamma > 0$.

It is also interesting to observe that $z \neq 0$ and $q/p = 1$ are valid in the PT-broken phase under consideration, which cannot be simultaneously true for the ordinary neutral-meson system. At this point, it is helpful to give some remarks on the CPT and CP symmetries in the neutral-meson system, and the ${\cal CPT}$ and ${\cal PT}$ symmetries in the PT-broken phase. Following the convention in Ref.~\cite{Bigi:2000yz}, one can write down the discrete space-time symmetry transformations for the neutral-meson system as
\begin{equation}
{\sf C}|P^0\rangle = -|\overline{P}^0\rangle \; , \quad {\sf P}|P^0\rangle = - |P^0\rangle \; , \quad {\sf T}|P^0\rangle = |P^0\rangle \; ,
\end{equation}
implying that ${\sf CP}|P^0\rangle = |\overline{P}^0\rangle$ and ${\sf CP}|\overline{P}^0\rangle = |P^0\rangle$. Observe that the time-reversal transformation will interchange the initial and final states, which form separately complete bases, and it is same in both the neutral-meson system and the PT-broken phase, i.e., ${\sf T} = {\cal T}$. In the matrix representation, we consider the two flavor eigenstates as $|P^0\rangle = (1, 0)^{\rm T}$ and $|\overline{P}^0\rangle = (0, 1)^{\rm T}$ and their Hermitian conjugated states $\langle P^0| = |P^0\rangle^\dagger = (1, 0)$ and $\langle \overline{P}^0| = |\overline{P}^0\rangle^\dagger = (0, 1)$. It is then straightforward to obtain
\begin{equation}
{\sf C} = \left(\begin{matrix} 0 & -1 \cr -1 & 0 \end{matrix}\right) \; , \quad {\sf P} = \left(\begin{matrix} -1 & 0 \cr 0 & -1 \end{matrix}\right) \; , \quad {\sf CP} = \left(\begin{matrix} 0 & 1 \cr 1 & 0 \end{matrix}\right) \; ,
\end{equation}
where one can observe that the matrix forms of the ${\sf CP}$ and ${\cal P}$ operators are exactly the same. It is not difficult to verify that the CPT or CP invariance in the neutral-meson system guarantees $M^{}_{11} = M^{}_{22}$ and $\Gamma^{}_{11} = \Gamma^{}_{22}$, while CP or T invariance leads to $\Im(M^{}_{12}) = \Im(\Gamma^{}_{12}) = 0$. However, as we have seen, the relations $M^{}_{11} = M^{}_{22}$, $\Gamma^{}_{11} \neq \Gamma^{}_{22}$, and $\Im(M^{}_{12}) = \Gamma^{}_{12} = 0$ hold in the PT-broken phase.

Since the transition probabilities for the flavor eigenstates in the neutral-meson system have been calculated, we can apply them directly to the PT-broken phase. Using Eq.~(\ref{eq:PP0P0}) as well as $\Delta m = 0$ and $\Gamma = 2\chi$, we find that
\begin{eqnarray}
{\cal P}^\prime_{aa}(t) &=& \left[\cosh^2(\gamma t) + \sinh^2(\gamma t) \frac{\cosh^2\alpha^\prime}{\sinh^2\alpha^\prime} + \sinh(2\gamma t) \frac{\cosh\alpha^\prime}{\sinh\alpha^\prime} \right] e^{-2\chi t} \nonumber \\
&=& \frac{\sinh^2(\alpha^\prime + \gamma t)}{\sinh^2\alpha^\prime}  e^{-2\chi t} \; ,
\label{eq:Paap}
\end{eqnarray}
and similarly using Eq.~(\ref{eq:PP0bP0}), we obtain
\begin{equation}
{\cal P}^\prime_{ab}(t) = \frac{1}{4} \left(e^{2\gamma t} + e^{-2\gamma t} - 2\right) e^{-2\chi t} \sqrt{(1 - \coth^2 \alpha^\prime)^2} = \frac{\sinh^2 (\gamma t)}{\sinh^2\alpha^\prime} e^{-2 \chi t}\; .
\label{eq:Pabp}
\end{equation}
Comparing the above results with the ones in Eqs.~(\ref{eq:Ppaa}) and (\ref{eq:Ppab}), we realize the additional exponential factor $e^{-2\chi t}$ and the exchange between the expressions of ${\cal P}^\prime_{aa}$ and ${\cal P}^\prime_{ab}$. Such an exchange can be understood by noticing the fact that the ${\cal PT}$-inner product and the ordinary inner product (i.e., the ${\cal T}$-inner product) differ by the parity operator ${\cal P}$ that causes the exchange of the final flavor eigenstates. In addition, in the limit of $t \to +\infty$, one can immediately verify that both ${\cal P}^\prime_{a a}(t)$ and ${\cal P}^\prime_{a b}(t)$ are proportional to $e^{-2(\chi - \gamma)t}$, which approaches zero due to the condition $\chi > \gamma$ for positive decay widths $\Gamma_1 > 0$ and $\Gamma_2 > 0$.

Before concluding this subsection, we make some helpful comments on the PT symmetry and the neutral-meson system. First, if one simply identifies the most general non-Hermitian Hamiltonian in Eq.~(\ref{eq:HMG}) with the PT-symmetric Hamiltonian in Eq.~(\ref{eq:H}) with $\phi = 0$, then it necessarily leads to $\Gamma^{}_1 < 0$ in the PT-broken phase. In such a case, the transition probabilities can be found by setting $\chi = 0$ in Eqs.~(\ref{eq:Paap}) and (\ref{eq:Pabp}). Consequently, the transition probabilities ${\cal P}^\prime_{aa}$ and ${\cal P}^\prime_{ab}$ become infinite in the limit of $t \to +\infty$. For this reason, it seems to be more interesting to consider the passive PT-symmetric Hamiltonian in Eq.~(\ref{eq:H_pass}), for which the masses and decay widths of the energy eigenstates are real and positive. Second, as the time evolution of the neutral-meson states $|P^0(t)\rangle$ and $|\overline{P}^0(t)\rangle$ is governed by the Schr\"{o}dinger equation with the Hamiltonian in Eq.~(\ref{eq:H_pass}), one can perform the gauge transformation~\cite{Ozdemir:2019}
\begin{eqnarray}
\left( \begin{matrix} |P^{\prime 0}(t)\rangle \cr |\overline{P}^{\prime 0}(t)\rangle \end{matrix}\right) = e^{+\chi t} \left( \begin{matrix} |P^0(t)\rangle \cr |\overline{P}^0(t)\rangle \end{matrix}\right)
\end{eqnarray}
such that the Schr\"{o}dinger equation for the gauge-transformed states $|P^{\prime 0}(t)\rangle$ and $|\overline{P}^{\prime 0}(t)\rangle$ is given by
\begin{eqnarray}
{\rm i}\frac{{\rm d}}{{\rm d}t} \left( \begin{matrix} |P^{\prime 0}(t)\rangle \cr |\overline{P}^{\prime 0}(t)\rangle \end{matrix}\right) = \left(\begin{matrix} \rho e^{+{\rm i}\varphi} & \sigma \cr  \sigma & \rho e^{-{\rm i}\varphi} \end{matrix}\right)  \left( \begin{matrix} |P^{\prime 0}(t)\rangle \cr |\overline{P}^{\prime 0}(t)\rangle \end{matrix}\right) \; .
\end{eqnarray}
Now, it is clear that the transition probabilities calculated in Subsection~\ref{sub: PTip} are applicable to the gauge-transformed states $|P^{\prime 0}(t)\rangle$ and $|\overline{P}^{\prime 0}(t)\rangle$, instead of the physical states $|P^0(t)\rangle$ and $|\overline{P}^0(t)\rangle$. Finally, it is worth pointing out that although only the transition amplitudes and probabilities in the framework of quantum mechanics are aimed for in the present work, the calculations can be performed in parallel for optical beam dynamics in PT-symmetric or PT-broken waveguides. In the latter case, exponential decay or amplification of the optical power takes place and has been experimentally observed~\cite{Feng:2017}.

\section{Summary and Conclusions}\label{sec: conclusion}

The basic properties of non-Hermitian Hamiltonians in both the PT-symmetric and PT-broken phases are interesting and their possible practical applications have recently received a lot of attention. In this work, we have focused on the flavor transitions in the two-level quantum system with PT-symmetric non-Hermitian Hamiltonians. Extending our previous investigation on the PT-symmetric phase with two real eigenvalues, we have considered the PT-broken phase, in which the two eigenvalues are complex conjugates to each other.

First, after solving the eigenvalues and eigenvectors of the non-Hermitian Hamiltonian in the PT-broken phase, we have explicitly constructed the charge-conjugation operator ${\cal C}$ and the metric operator $\eta^\prime$, for which the identities ${\cal C} = \mathbb{1}_2$ and $\eta^\prime = {\cal P}$ are valid. Second, using the ${\cal PT}$-inner product, we have calculated the transition amplitudes and probabilities for the flavor eigenstates, i.e., $|u^\prime_\alpha\rangle \to |u^\prime_\beta\rangle$ for $\alpha, \beta = a, b$. After introducing the ${\cal CPT}$ flavor eigenstates ${\cal CPT}|\tilde{u}^\prime_a\rangle = +|\tilde{u}^\prime_a\rangle$ and ${\cal CPT}|\tilde{u}^\prime_b\rangle = -|\tilde{u}^\prime_b\rangle$ as the final states, we have found that the difference $\tilde{\cal P}^\prime_{aa} - \tilde{\cal P}^\prime_{ab}$ between, instead of the sum $\tilde{\cal P}^\prime_{aa} + \tilde{\cal P}^\prime_{ab}$ of, the corresponding transition probabilities, vanishes and is time-independent. However, the probabilities themselves in the PT-broken phase have been found to be infinite in the limit of $t \to +\infty$, which is totally different from the corresponding result in the PT-symmetric phase. Third, in analogy to the neutral-meson system, we have also calculated the transition probabilities using the ordinary inner product, which is equivalent to the ${\cal T}$-inner product, and observed that the infinite-time behavior of the probabilities originates from the negative decay width $\Gamma^{}_1 = -2\gamma < 0$ of one energy eigenstate in the PT-broken phase. We also make a connection between the neutral-meson system and the passive PT-symmetric Hamiltonian, where an extra global decay term (described by the parameter $\chi > 0$) is present, and thus, $\Gamma^{}_1 = 2(\chi - \gamma) > 0$ and $\Gamma^{}_2 = 2(\chi + \gamma) > 0$. Even in this case, the relations $M^{}_{11} = M^{}_{22}$, $\Gamma^{}_{11} \neq \Gamma^{}_{22}$, and $\Im(M^{}_{12}) = \Gamma^{}_{12} = 0$ are different from those accessible in the ordinary neutral-meson system. For this reason, since the PT-broken phase cannot be used to describe the neutral-meson system, one might have to find practical applications of the PT-broken phase in dynamical systems beyond particle physics. In addition, we have presented a discussion on the exceptional point for several different and interesting cases of non-Hermitian Hamiltonians (including the passive PT-symmetric Hamiltonian with a global exponential decay or amplification) that have applications in various physical systems such as photonics, phononics, and other open quantum and PT-symmetric systems.

Finally, the results that have been presented in this work indicate that the PT-broken phase and the exceptional point have very different properties compared with the PT-symmetric phase and this deserves further exploration. As shown in Refs.~\cite{Wiersig1, Wiersig2, Sensors}, the microcavity sensors prepared at the exceptional point will be much more sensitive to small perturbations, which can be implemented to realize a one-particle detection. In a similar way, the practical applications of the non-Hermitian Hamiltonian in the PT-broken phase may be accomplished only after coupling it to another system. This is the case for the neutral-meson system, where the weak interaction is switched on in order for the neutral mesons to decay. As we have mentioned, realistic applications may be lying beyond particle physics. We leave all these important points for future works.

\section*{Acknowledgments}

T.O.~acknowledges support by the Swedish Research Council (Vetenskapsr{\aa}det) through contract No.~2017-03934 and the KTH Royal Institute of Technology for a sabbatical period at the Uni\-versity of Iceland. The work of S.Z.~was supported in part by the National Natural Science Foundation of China under grant No.~11775232 and No.~11835013, and by the CAS Center for Excellence in Particle Physics.

S.Z.~is also greatly indebted to his family, friends and colleagues for all their support and encouragement during his isolated stay in Huanggang, Hubei, China, where the new corona virus is particularly prevalent and the present work is completed.

\section*{Data Availability}

The data that support the findings of this study are available from the corresponding author upon reasonable request.

\end{document}